\documentclass[final,5p,times,twocolumn]{elsarticle}

\usepackage{amssymb}

\usepackage{lineno}

\begin{document}
\begin{frontmatter}


\title{Refining the evaluation of eigen emittances measured by the dedicated four-dimensional emittance scanner ROSE}


\author{C.~Xiao, X.N.~Du, L.~Groening, and M.~Maier}

\address{GSI Helmholtzzentrum f\"ur Schwerionenforschung GmbH, D-64291 Darmstadt, Germany}

\begin{abstract}
A dedicated device to fully determine the four-dimensional beam matrix, called ROSE (ROtating System for Emittance measurements) was successfully commissioned. Results obtained with $^{83}$Kr$^{13+}$ at 1.4~MeV/u are reported in Phys. Rev. Accel. Beams {\bf19}, 072802 (2016). Coupled moments were determined with an accuracy of about~10\%, which is sufficiently low to reliably determine a lattice which could decouple the beam. However, the remaining uncertainty on the corresponding eigen emittances was still considerable high. The present paper reports on improvement of the evaluation procedure which lowers the inaccuracy of measured eigen emittances significantly to the percent level. The method is based on trimming directly measured data within their intrinsic measurement resolution such that the finally resulting quantity is determined with high precision. 
\end{abstract}


\end{frontmatter}


\section{Introduction}
Projected emittances are figures of merit for each accelerator, hence their reduction, preferably without beam loss, is of fundamental interest. This reduction can be accomplished through elimination of eventual inter-plane correlations, i.e., coupling between different planes of the six-dimensional (6d) phase space. However, in order to do so, the correlations must be known quantitatively. If these are available to sufficient accuracy, a dedicated decoupling beam line can be designed to reduce the projected emittances. For ion energies up to about 150~keV/u pepper pots can be used~\cite{Kondrashev,Kremers:DIPAC07-TUPC24,Nagatomo} for 4d diagnostics but at higher energies these devices suffer from fluorescence effects on the screen and results are not reliable~\cite{Peter}.

To overcome this limitation with respect to applicable ion energy, coupled moments were measured through scanning skewed quadrupoles followed by a regular slit-grid emittance scanner~\cite{Chen1}. The 4d measurement time was considerable reduced by developing the dedicated device ROSE (ROtating System for Emittance measurements)~\cite{Chen2}, allowing to perform full 4d rms-emittance measurements of ion beams independent of their energy and time structure. It has been designed, built, and successfully commissioned at the UNILAC~\cite{Groening:IPAC16-MOPOY017} of GSI. In 2018 the device and the underlying data evaluation method were patented~\cite{patent}. Currently, technology transfer to industry is 
ongoing~\cite{Maier:IBIC19-THAO03,MMM}.

ROSE is a standard single plane slit-grid emittance scanner which can be rotated around the beam axis. Slit and grid are installed inside of a rotatable vacuum chamber. This chamber does not rotate during the emittance measurements itself. Rather emittance scans are performed at different rotation angles of the chamber combined with two different beam optics in front of the chamber. The combination of emittance scans at different angles and different optics provides quantitative access to the rms inter-plane correlations, hence paving the path towards their elimination.

During successful commissioning of ROSE it was found that the inter-plane correlations and the optics required to remove them were determined with sufficient precision. However, the resulting eigen emittances are still prone to relatively large relative errors of several~10\%. In order to address this issue the data evaluation analysis has been revisited~\cite{MMM}. A new method basing on trimming directly measured quantities within their measurement precision has been elaborated. It aimed for determination of a trimmed set of measured data that delivers the most consistent set of identical quantities being derived in different ways from the trimmed set. The re-evaluated eigen emittances have significantly lower errors and are still within the error bars being derived using the initial method reported in~\cite{Chen2}.

After an introduction of rms parameters which quantify 4d particle distributions, the third section briefly recapitulates the previously applied evaluation algorithm of ROSE~\cite{Chen2} and the corresponding results. The fourth section describes the newly elaborated evaluation, where the previous algorithm is split into many procedures of same kind using different input, and presents the corresponding refined results. Finally, ROSE's capability to provide for proper determination of a decoupling lattice is confirmed for the new evaluation method. The paper closes with a discussion of the theoretical base of the new evaluation method, its current limitations and future work to be done.
\section{Rms quantities of four dimensional distributions}
The full transverse 4d symmetric beam matrix, made from the second moments, contains ten independent elements
\begin{equation}
C=
\left(
\begin{array}{clcr}
\langle xx \rangle &  \langle xx'\rangle &  \langle xy\rangle & \langle xy'\rangle \\
\langle x'x\rangle &  \langle x'x'\rangle & \langle x'y\rangle & \langle x'y'\rangle \\
\langle yx\rangle &  \langle yx'\rangle &  \langle yy\rangle & \langle yy'\rangle \\
\langle y'x\rangle &  \langle y'x'\rangle & \langle y'y\rangle & \langle y'y'\rangle\\
\end{array}
\right)\,,
\end{equation}

of which four of them quantify the inter-plane correlations. If one of these four elements is different from zero, the beam is $x$-$y$ coupled. Usually just separated measurements in the $x$-$x'$ and $y$-$y'$ sub phase-planes are performed, and the projected beam rms-emittances $\varepsilon_{x,y}$ are the square roots of the determinants of the on-diagonal sub-matrices. These projected rms-emittances can be considerably enlarged by (usually) unwanted inter-plane correlations, and these may be created and/or augmented in solenoids and tilted quadrupoles and dipoles. They can also be created by simple beam loss, if for instance the upper right part of a misaligned beam is lost somewhere.

Diagonalization of the 4d beam matrix yields the eigen emittances $\varepsilon_{1,2}$ as~\cite{Chen3}
\begin{equation}
\label{eq22}
{\varepsilon_{1,2}}=\frac{1}{2} \sqrt{-tr[(CJ)^2] \pm \sqrt{tr^2[(CJ)^2]-16 \left| C \right|}}\,,
\end{equation}
where $J$ is 
\begin{equation}
J:=
\left(
\begin{array}{clcr}
0 &  1 &  0 & 0 \\
-1 &  0 &  0 & 0 \\
0 &  0 &  0 & 1 \\
0 &  0 & -1 & 0 \\
\end{array}
\right)\,.
\end{equation}

Eigen emittances are invariant under symplectic transformations, and the coupling parameter $t$ quantifies the inter-plane coupling through
\begin{equation}
\label{t value}
t:=\frac{\varepsilon_x  \varepsilon_y}{\varepsilon_1  \varepsilon_2}-1 \geqslant 0
\end{equation}
and if $t$ is equal to zero, there are no inter-plane correlations and the projected rms-emittances are equal to the eigen emittances.
\section{Previous evaluation method}
\begin{figure}[hbt]
\centering
\includegraphics*[width=80mm,clip=]{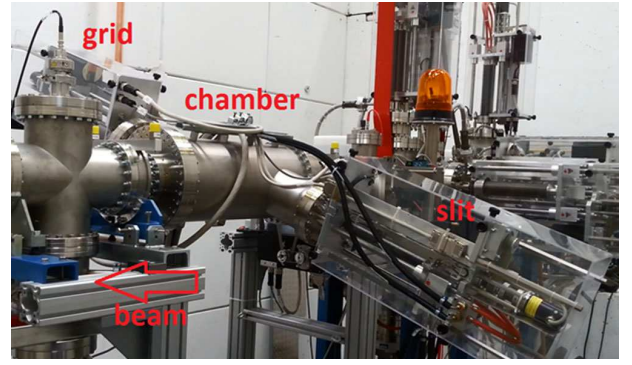}
\caption{ROtating System for Emittance measurements (ROSE): a single plane slit and grid emittance measurement device housed in a chamber which can be rotated around the beam axis.}
\label{ROSE}
\end{figure}

The ROSE scanner shown in Fig.~\ref{ROSE} comprises a slit followed by a grid installed in one chamber. At the location of the slit each emittance measurement provides three measured beam moments which are $\langle rr \rangle$, $\langle rr' \rangle$, and $\langle r'r' \rangle$. Depending on the rotation angle $\theta$ of the vacuum chamber $r$ corresponds to certain directions, i.e., $\theta $=0$^{\circ}$ to the horizontal ($x$) and $\theta $=90$^{\circ}$ to the vertical ($y$) direction.

\begin{figure}[hbt]
\centering
\includegraphics*[width=80mm,clip=]{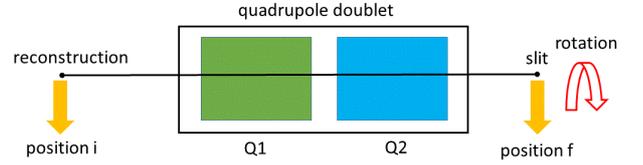}
\caption{Sketch of the beam line comprising a quadrupole doublet and ROSE. Beam moments are measured at the ROSE slit and the 4d beam matrix is finally determined at the reconstruction position.}
\label{doublet}
\end{figure}

The position at the beam line, where the 4d beam matrix shall be finally determined, is located before ROSE itself and will be called the reconstruction position in the following. This reconstruction position and ROSE are separated by a non-coupling regular quadrupole doublet as depicted in~Fig.~\ref{doublet}. Two different settings $a$ and $b$ of the doublet are applied and for each of them ROSE measures at three different angles $\theta$=0$^{\circ}$, $\theta$=90$^{\circ}$, and $\theta$=$\Theta$$^{\circ}$, (any angle which is not equivalent to 0$^{\circ}$ or 90$^{\circ}$) providing a set of \mbox{2$\times$3$\times$3=18} measured second beam moments.

The searched off-diagonal matrix elements at the reconstruction position can be implicitly expressed through an over-determined set of linear equations as
\begin{equation}
\label{coupling001}
\Gamma_{11,41}\langle xy \rangle+\Gamma_{12,42}\langle xy' \rangle+\Gamma_{13,43}\langle x'y \rangle+\Gamma_{14,44}\langle x'y' \rangle=\Lambda_{1,4}
\end{equation}
\begin{equation}
\label{coupling002}
\Gamma_{21,51}\langle xy \rangle+\Gamma_{22,52}\langle xy' \rangle+\Gamma_{23,53}\langle x'y \rangle+\Gamma_{24,54}\langle x'y' \rangle=\Lambda_{2,5}
\end{equation}
\begin{equation}
\label{coupling006}
\Gamma_{31,61}\langle xy \rangle+\Gamma_{32,62}\langle xy' \rangle+\Gamma_{33,63}\langle x'y \rangle+\Gamma_{34,64}\langle x'y' \rangle=\Lambda_{3,6}\,,
\end{equation}
where the $\Gamma$ matrix is made from the beam transport matrix elements given by the two quadrupole settings $a$ and~$b$:
\begin{equation}
\label{a}
\Gamma_{11}=m^a_{11}m^a_{33},~\Gamma_{12}=m^a_{11}m^a_{34},~\Gamma_{13}=m^a_{12}m^a_{33},~\Gamma_{14}=m^a_{12}m^a_{34}
\end{equation}
\begin{equation}
\Gamma_{21}=m^a_{11}m^a_{43}+m^a_{21}m^a_{33},~\Gamma_{22}=m^a_{11}m^a_{44}+m^a_{21}m^a_{34}
\end{equation}
\begin{equation}
\Gamma_{23}=m^a_{12}m^a_{43}+m^a_{22}m^a_{33},~\Gamma_{24}=m^a_{12}m^a_{44}+m^a_{22}m^a_{34}
\end{equation}
\begin{equation}
\Gamma_{31}=m^a_{21}m^a_{43},~\Gamma_{32}=m^a_{21}m^a_{44},~\Gamma_{33}=m^a_{22}m^a_{43},~\Gamma_{34}=m^a_{22}m^a_{44}
\end{equation}
\begin{equation}
\Gamma_{41}=m^b_{11}m^b_{33},~\Gamma_{42}=m^b_{11}m^b_{34},~\Gamma_{43}=m^b_{12}m^b_{33},~\Gamma_{44}=m^b_{12}m^b_{34}
\end{equation}
\begin{equation}
\Gamma_{51}=m^b_{11}m^b_{43}+m^b_{21}m^b_{33},~\Gamma_{52}=m^b_{11}m^b_{44}+m^b_{21}m^b_{34}
\end{equation}
\begin{equation}
\Gamma_{53}=m^b_{12}m^b_{43}+m^b_{22}m^b_{33},~\Gamma_{54}=m^b_{12}m^b_{44}+m^b_{22}m^b_{34}
\end{equation}
\begin{equation}
\Gamma_{61}=m^b_{21}m^b_{43},~\Gamma_{62}=m^b_{21}m^b_{44},~\Gamma_{63}=m^b_{22}m^b_{43},~\Gamma_{64}=m^b_{22}m^b_{44}\,,
\end{equation}
and the vector $\Lambda$ is constructed from the 18 beam moments directly measured at the slit of ROSE
\begin{equation}
\label{Gamma_ab}
\Lambda_{1,4}=\langle xy \rangle^{a,b}=\frac{\langle rr \rangle^{a,b}_{\theta_2}-\cos^2\theta_2\langle rr \rangle^{a,b}_{\theta_1}-\sin^2\theta_2\langle rr \rangle^{a,b}_{\theta_3}}{2\sin\theta_2\cos\theta_2}
\end{equation}
\begin{equation}
\Lambda_{2,5}=\langle xy' \rangle^{a,b}+\langle x'y \rangle^{a,b}=\frac{\langle rr' \rangle^{a,b}_{\theta_2}-\cos^2\theta_2\langle rr' \rangle^{a,b}_{\theta_1}-\sin^2\theta_2\langle rr' \rangle^{a,b}_{\theta_3}}{\sin\theta_2\cos\theta_2}
\end{equation}
\begin{equation}
\Lambda_{3,6}=\langle x'y' \rangle^{a,b}=\frac{\langle r'r' \rangle^{a,b}_{\theta_2}-\cos^2\theta_2\langle r'r' \rangle^{a,b}_{\theta_1}-\sin^2\theta_2\langle r'r' \rangle^{a,b}_{\theta_3}}{2\sin\theta_2\cos\theta_2}\,,
\end{equation}
$\theta_1$=0$^{\circ}$, $\theta_3$=90$^{\circ}$, and $\theta_2$=$\Theta$$^{\circ}$ indicate the respective rotation angle and $a$ and $b$ indicate the applied quadrupole setting. For this over-determined system of equations a unique solution at the reconstruction position can be defined as 
\begin{equation}
\label{gamma_solution}
\left(
\begin{array}{clcr}
\langle xy \rangle \\
\langle xy' \rangle \\
\langle x'y \rangle \\
\langle x'y' \rangle \\
\end{array}
\right)
=\left(\Gamma^T \Gamma\right)^{-1} \Gamma^T \Lambda\,.
\end{equation}

The 18 directly measured beam moments at the slit inhabit intrinsic measurement errors of about~10\%, and these will enter into the final results for the off-diagonal moments at the reconstruction position. The condition number $\kappa\left(\Gamma\right)$, defined through a Frobenius from~\cite{Chen2}, quantifies how much the output of a function can change for a small change of its argument:
\begin{equation}
\label{cond}
\kappa\left(\Gamma\right):={\Vert \Gamma \Vert} {\Vert \Gamma^{\dagger} \Vert}\,,
\end{equation}
with
\begin{equation}
\label{pseudo-inverse}
\Gamma^{\dagger}=
\left(\Gamma^T \Gamma\right)^{-1} \Gamma^T\,.
\end{equation}

The off-diagonal correlated second moments are less sensitive to moment measurement errors if $\kappa\left(\Gamma\right)$ is small. If the $\Gamma$-matrix is well-conditioned and the input data forming the $\Lambda$-vector is sufficiently precise, the final solution for the four coupled moments will be accurate. Hence, the accuracy of the coupled second moments measurement depends on $\kappa\left(\Gamma\right)$ and on the precision of the $\Lambda$-vector.

In front of ROSE a skew quadrupole triplet was applied to create considerable inter-plane coupling. The two doublet settings $a$ and $b$ used for the measurements are listed in Tab.~\ref{tab_0}. The respective ROSE rotation angles and the Twiss parameters calculated from the measured beam moments are listed in Tab.~\ref{tab_5}. The latter were calculated from the moments as
\begin{equation}
\label{Twiss_Moments}
\varepsilon_{rms}:=\sqrt{\langle rr \rangle\langle r'r' \rangle-{\langle rr' \rangle}^2},~~
\alpha_{rms}:=-\frac{\langle rr' \rangle}{\varepsilon},~~\beta_{rms}:=\frac{\langle rr \rangle}{\varepsilon}\,.
\end{equation} 
\begin{table}
\caption{\label{tab_0} Quadrupole doublet settings $a$ and $b$ providing a small condition number of~$\kappa$=13.14 together with full transmission. Positive signs mean horizontal focusing.}
\centering
\begin{tabular}{c|c|c}
\hline
\hline
setting & first quadrupole [T/m] & second quadrupole [T/m]\\
\hline
a & 9.38 &-10.19\\
b & 13.16 &-12.57\\
\hline
\hline
\end{tabular}
\end{table}
\begin{table}
\caption{\label{tab_5} Measured projected rms-emittances and Twiss parameters at rotation angles $\theta_1$=0$^{\circ}$, $\theta_2$=30$^{\circ}$, and $\theta_3$=90$^{\circ}$ using settings $a$ and $b$.}
\centering
\begin{tabular}{c|c|c|c|c}
\hline
\hline
rotation& setting &$\alpha_{rms}$ & $\beta_{rms}$ [m/rad]& $\varepsilon_{rms}$ [mm~mrad]\\
\hline
0$^{\circ}$ & a &-0.135& 4.611 & 3.148\\
0$^{\circ}$ & b &0.100 & 3.950 & 3.125\\
30$^{\circ}$& a &-0.547 &2.254 &3.200\\
30$^{\circ}$& b &-0.802 &2.677 &4.673\\
90$^{\circ}$& a &-2.452 &8.790 & 3.405\\
90$^{\circ}$& b &-2.690 &7.350 & 3.330\\
\hline
\hline
\end{tabular}
\end{table}

\begin{figure}[hbt]
\centering
\includegraphics*[width=80mm,clip=]{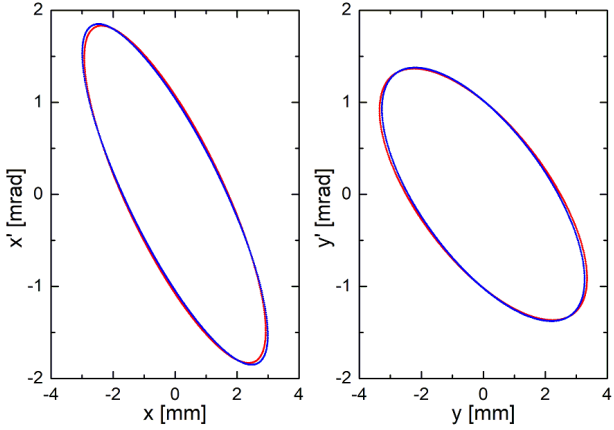}
\caption{Projected rms-ellipses in horizontal and vertical planes obtained from measurements with the two setting $a$ and $b$ and back transformation to the reconstruction position.}
\label{check}
\end{figure}

The two projected rms-ellipses at rotation angles $\theta_1$ and $\theta_3$ using settings $a$ and $b$ were transformed back to the reconstruction position and are compared to each other in~Fig.~\ref{check}.
The overlap is excellent in both planes, hence the uncorrelated moments at the reconstruction position are determined with high reliability.

\begin{figure}[hbt]
\centering
\includegraphics*[width=80mm,clip=]{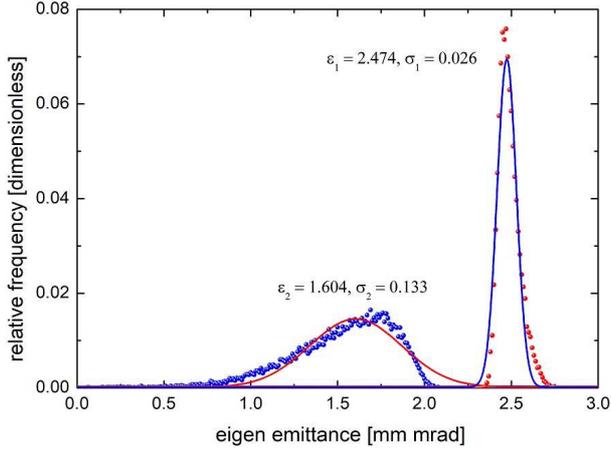}
\caption{Relative frequencies of the eigen emittances with $\pm$1\% random measured second moment errors. Red and blue dots indicate the eigen emittances and blue and red solid lines indicate the fits using Gaussian functions. Applying error analysis, the mean eigen emittances are obtained as $\varepsilon_1$=2.474$\pm0.026$~mm~mrad and $\varepsilon_2$=1.604$\pm$0.133~mm~mrad.}
\label{error_analysis}
\end{figure}

Plugging all measured moments listed in Tab.~\ref{tab_5} into Eq.~(\ref{gamma_solution}) delivers the complete beam moments matrix (in units of mm and mrad)
\begin{equation}
\label{C_prab}
C=
\left(
\begin{array}{clcr}
+8.5713 &  -4.3421 & -3.8329    & -1.1485 \\
-4.3421  &  +3.3555 &  -0.5428  & +1.5181 \\
-3.8329   &  -0.5428   & +11.2017 & -3.0531\\
 -1.1485  & +1.5181 &   -3.0531  & +1.8672 \\
\end{array}
\right)\,,
\end{equation}
at the reconstruction position. Evaluation of the eigen emittances from matrix $C$ gives $\varepsilon_1$=2.472~mm~mrad, $\varepsilon_2$=1.582~mm~mrad, and $t$=1.74. These results have been reported in~\cite{Chen2} together with the observation that the second moments, including the coupled ones, are determined with good precision. However, the obtained precision of the corresponding eigen emittances is significantly less. The error study revealed that especially the lower of the eigen emittances is subject to strong variations even for very small changes of the measured second moments. This shall be illustrated here by Fig.~\ref{error_analysis}: the eigen emittances were evaluated from solving~Eq.~(\ref{eq22}). This was done for many different sets of $\Lambda$ vectors and the sets were generated by random variation of the 18 directly measured moments. Accordingly the two eigen emittances were broadened to a spectrum each. These spectra, depicted in Fig.~\ref{error_analysis}, were created by the random variation of the measured moments within just~$\pm$1\% around the measured value. Albeit this tiny variation the full spectrum width of the small eigen emittance is about~$\pm$30\% of its mean value, i.e., a magnification of the error by a factor of~30.

In the special case considered here there are $k$=4 unknown coupled second moments and $n$=6 linear equations in~Eq.~(\ref{coupling001}) to~Eq.~(\ref{coupling006}). At least $k$=4 of them must be selected to obtain a properly defined solution. There are
\begin{equation}
\label{zhuhe}
C_n^k=\frac{n!}{k!\left(n-k\right)!}=15
\end{equation}
possible selections delivering one set $j$ of solutions each:
\begin{equation}
\label{15}
\left(
\begin{array}{clcr}
{\langle xy \rangle} \\
{\langle xy' \rangle}\\
{\langle x'y \rangle}\\
{\langle x'y' \rangle} \\
\end{array}
\right)_j
=
\Gamma_j^{-1}\Lambda_j,~~~j=1,2,\cdots 15
\end{equation}
as
\begin{equation}
\label{coupling011}
\left(
\begin{array}{clcr}
\langle xy \rangle \\
\langle xy' \rangle\\
\langle x'y \rangle\\
\langle x'y' \rangle \\
\end{array}
\right)_1
=
\left(
\begin{array}{clcr}
\Gamma_{11} & \Gamma_{12} & \Gamma_{13} & \Gamma_{14} \\
\Gamma_{21} & \Gamma_{22} & \Gamma_{23} & \Gamma_{24} \\
\Gamma_{31} & \Gamma_{32} & \Gamma_{33} & \Gamma_{34} \\
\Gamma_{41} & \Gamma_{42} & \Gamma_{43} & \Gamma_{44} \\
\end{array}
\right)^{-1}
\left(
\begin{array}{clcr}
\Lambda_1 \\
\Lambda_2 \\
\Lambda_3 \\
\Lambda_4
\end{array}
\right)\,,
\end{equation}
\begin{equation}
\cdots
\end{equation}
\begin{equation}
\label{coupling012}
\left(
\begin{array}{clcr}
\langle xy \rangle \\
\langle xy' \rangle\\
\langle x'y \rangle\\
\langle x'y' \rangle \\
\end{array}
\right)_{15}
=
\left(
\begin{array}{clcr}
\Gamma_{41} & \Gamma_{42} & \Gamma_{43} & \Gamma_{44} \\
\Gamma_{51} & \Gamma_{52} & \Gamma_{53} & \Gamma_{54} \\
\Gamma_{61} & \Gamma_{62} & \Gamma_{63} & \Gamma_{64} \\
\Gamma_{31} & \Gamma_{32} & \Gamma_{33} & \Gamma_{34} \\
\end{array}
\right)^{-1}
\left(
\begin{array}{clcr}
\Lambda_4 \\
\Lambda_5 \\
\Lambda_6 \\
\Lambda_3 \\
\end{array}
\right)\,.
\end{equation}

The square matrix $\Gamma_j$, made from beam transfer matrix elements, is considered as ill-conditioned if its condition number is infinitely large. In turn, if all 15 condition numbers are sufficiently small, all possible selections should produce similar correlated second moments, i.e., the spread among the 15 solutions should be reasonably low. For the slit-grid emittance measurements of ROSE the finite slit and grid resolutions together with background noise cause errors of about $\pm$10\% of the 18 directly measured moments.

\begin{figure}[hbt]
\centering
\includegraphics*[width=80mm,clip=]{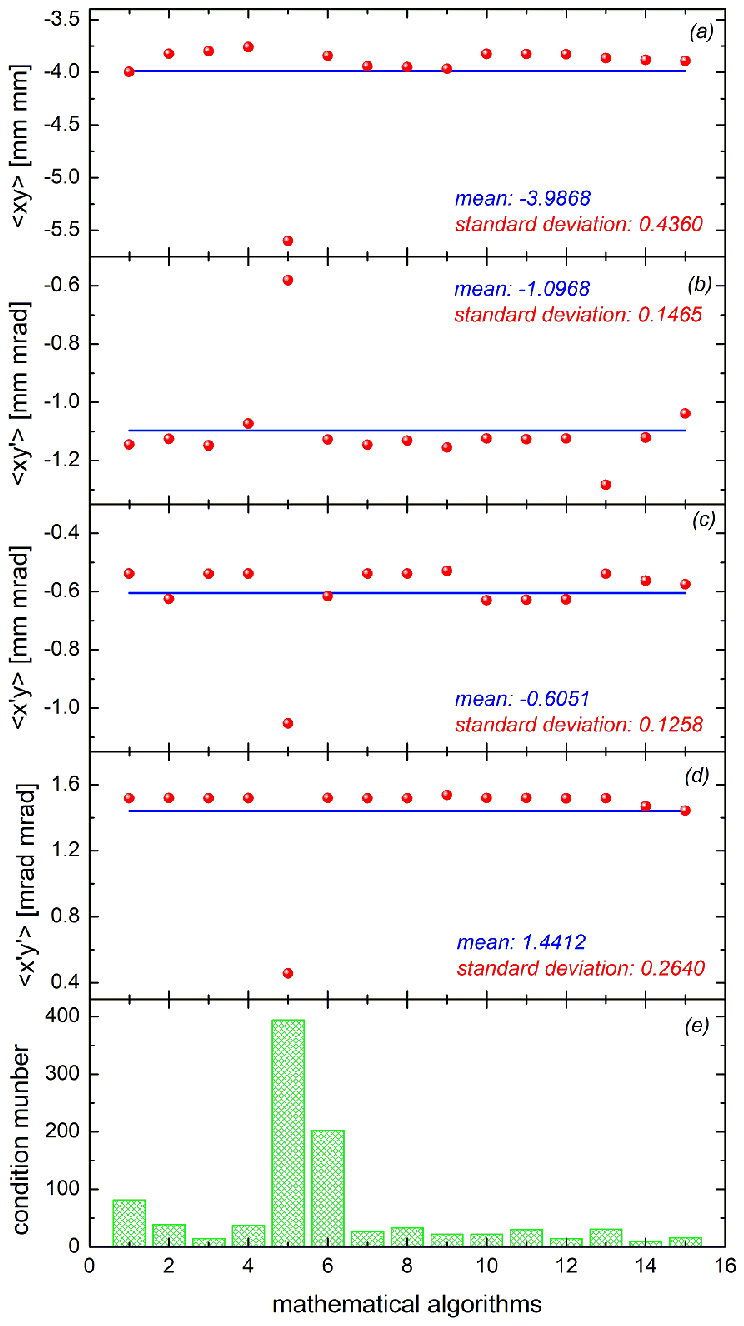}
\caption{Coupled beam moments evaluated from Tab.~\ref{tab_5} using all 15 possible selections. Dots indicate individual moments and solid lines indicate their means. (a) to (d): individual coupled beam moments and their means. (e): condition numbers.}
\label{correlated_moments_15}
\end{figure}

\begin{figure}[hbt]
\centering
\includegraphics*[width=80mm,clip=]{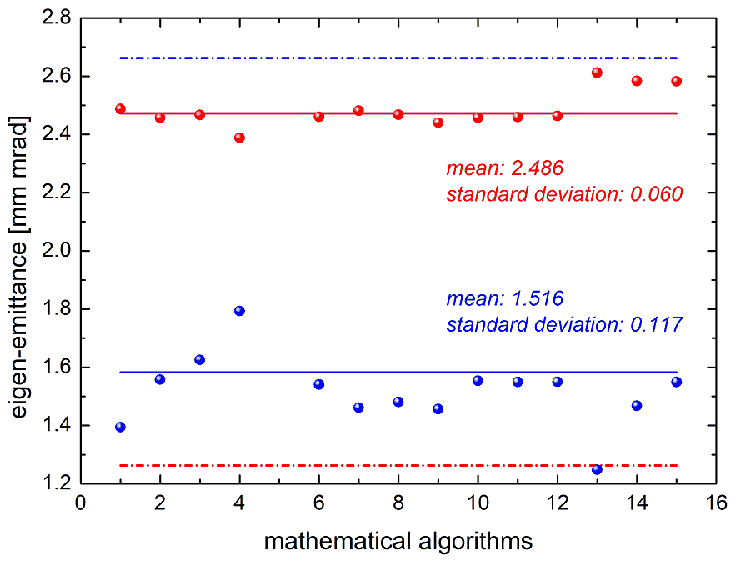}
\caption{Eigen emittances evaluated from Tab.~\ref{tab_5} using all 15 possible selections. Dots indicate individual eigen emittances and solid lines indicate their means. Dashed lines indicate the eigen emittances obtained from matrix $C^\star$. The data of mathematical algorithm$\#$5 are removed due to the unreliable evaluation results: $\varepsilon_1$=4.800~mm~mrad (real number) and $\varepsilon_2$=2.807$i$~mm~mrad (complex number).}
\label{eigen_emittances_1}
\end{figure}

In the following coupled moments and eigen emittances are evaluated from applying~Eq.~(\ref{15}) to all 15 possible selections to pick four out of six equations from~Eq.~(\ref{coupling001}) to~Eq.~(\ref{coupling006}). The results from these 15 selections are plotted in Fig.~\ref{correlated_moments_15} and Fig.~\ref{eigen_emittances_1}. The mean moments matrix
\begin{equation}
\label{test}
C^\star\,:=\,\frac{1}{15}\sum_{j=1}^{15}C_j
\end{equation}
from all 15 selections is (in units of mm and mrad)
\begin{equation}
\label{C_chen}
C^\star=
\left(
\begin{array}{clcr}
+8.5713 &  -4.3421 & -3.9868    & -1.0968 \\
-4.3421  &  +3.3555 &  -0.6051  & +1.4412 \\
-3.9868   &  -0.6051   & +11.2017 & -3.0531\\
-1.0968   & +1.4412 &   -3.0531  & +1.8672 \\
\end{array}
\right)\,.
\end{equation}

Evaluation of the eigen emittances from matrix $C^\star$ results in $\varepsilon_1$=2.662~mm~mrad, $\varepsilon_2$=1.263~mm~mrad, and $t$=2.19. Comparing matrices $C$ and $C^\star$ shows that the coupled moments are quite similar but the corresponding eigen emittances, and hence the $t$-values, differ considerably. 

Figures~\ref{correlated_moments_15} and~\ref{eigen_emittances_1} show that each selection delivers its individual values for the beam moments and eigen emittances. Very strictly speaking these 15 sets of results contradict to each other. This is from the fact that the input for the 15 evaluations from Tab.~\ref{tab_5} is subject to intrinsic inaccuracies from the measurements, hence the input itself is already intrinsically inconsistent. Taking this into account one cannot expect that all 15 selections deliver identical results. The inconsistency of the input transfers into an inconsistency of the final result. The new evaluation method approaches this fact in the way that it trims the input within its intrinsic inaccuracy such that all 15 selections deliver practically identical results. In turn 15 consistent final results indicate that the corresponding (trimmed) input is consistent itself.
\section{Improved evaluation method}
The previous method aimed for finding a best fitting solution to the over-determined system of linear equations. As input served the 18 directly measured beam moments at the ROSE slit and the known transport matrix elements comprising $\Gamma _j$ of Eq.~(\ref{15}). The final solution for the coupled moments was determined such that it fits best to all six sub-equations of~Eq.~(\ref{coupling001}) to~Eq.~(\ref{coupling006}). The concept of this approach was implicitly to assume that the input is correct (albeit being aware that it is subject to measurement errors) and to accept that the final solution is just a compromise of matching best all sub-equations of~Eq.~(\ref{coupling001}) to~Eq.~(\ref{coupling006}). Accordingly, this approach acknowledges that the final solution has an error as it does not strictly fulfill all sub-equations simultaneously. 

Figure~\ref{correlated_moments_15} shows the results for the coupled moments obtained from the 15 possible selections. These results are not identical but show spreads with given standard deviations, which are stated in the figure's legend. The new evaluation method trims the 18 directly measured moments such that these standard deviations are minimized. It does explicitly not trim the transport matrix elements contained in $\Gamma $ since the relative accuracy of these is at least two orders of magnitude better compared to the accuracy of the measured beam moments. The properties of the elements forming the beam line corresponding to $\Gamma $ are known to very high precision from rigorous quadrupole magnet field mapping and alignment through laser tracking.

The improved method inverts the initial assumptions and the way to interprete the final result. Instead of assuming fixed and correct input delivering a result contradicting to the equations it results from, the new evaluation method implicitly accounts for the intrinsic uncertainty of the input itself within the resolution of the measurements. The new method varies the input parameters within this resolution but claims that the final result shall meet all sub-equations of~Eq.~(\ref{coupling001}) to~Eq.~(\ref{coupling006}). Hence, the final result is fully consistent to all equations it is derived from. It is not a compromise just fitting best all these equations. Instead it is consistent with all sub-equations.

A routine implemented with MATHCAD~\cite{Mathcad} has been developed to trim the 18 measured moments in order to minimize the standard deviations of the reconstructed coupled moments. A set of functions of the coupled moments depending on the directly measured moments is defined as
\begin{equation}
\label{mathcad21}
\left(
\begin{array}{clcr}
\langle xy \rangle\\
\langle xy' \rangle\\
\langle x'y \rangle\\
\langle x'y' \rangle
\end{array}
\right)_j
=\digamma_j\left(\zeta^{a,b}_{\theta_1,\theta_2,\theta_3}\right)=
\left(
\begin{array}{clcr}
\Xi\left(\zeta^{a,b}_{\theta_1,\theta_2,\theta_3}\right)\\
\Pi\left(\zeta^{a,b}_{\theta_1,\theta_2,\theta_3}\right)\\
\Upsilon\left(\zeta^{a,b}_{\theta_1,\theta_2,\theta_3}\right)\\
\Omega\left(\zeta^{a,b}_{\theta_1,\theta_2,\theta_3}\right)
\end{array}
\right)_j\,,
\end{equation}
where $j$ refers to one of the 15 selections and $\zeta^{a,b}_{\theta_1,\theta_2,\theta_3}$ stands for the 18 directly measured moments to be trimmed
\begin{equation}
\zeta^{a,b}_{\theta_1,\theta_2,\theta_3}=\left(\varepsilon,\alpha,\beta\right)^{a,b}_{\theta_1,\theta_2,\theta_3}=\left(\varepsilon,\alpha,\beta\right)^a_{\theta_1},\left(\varepsilon,\alpha,\beta\right)^a_{\theta_2}\cdots\left(\varepsilon,\alpha,\beta\right)^b_{\theta_3}\,.
\end{equation}
The standard deviation of the coupled second moment $\mu$ at the reconstruction position is
\begin{equation}
\label{mathcad22}
\sigma^{\mu}=\sqrt {\frac{1}{n}\sum_{p=1}^{n} \left[\mu_p-\frac{1}{n}\sum_{q=1}^{n} \mu_q \right]^2}~~~~~~n=15\,.
\end{equation}

Applying the second moment $\langle xy \rangle$ for instance, the standard deviation of the coupled second moment $\sigma^{\langle xy \rangle}$ is
\begin{equation}
\sigma^{\langle xy \rangle}:=\sqrt {\frac{1}{n}\sum_{p=1}^{n} \left\{\Xi_p\left[\left(\varepsilon,\alpha,\beta\right)^{a,b}_{\theta_1,\theta_2,\theta_3}\right]-\frac{1}{n}\sum_{q=1}^{n} \Xi_q\left[\left(\varepsilon,\alpha,\beta\right)^{a,b}_{\theta_1,\theta_2,\theta_3}\right] \right\}^2}\,.
\end{equation}

The optimization function $\Re^{\langle xy \rangle}$ basing on the Mathcad program (KNITRO solving algorithm) has been adopted to seek the values of $\left(\varepsilon,\alpha,\beta\right)^{a,b}_{\theta_1,\theta_2,\theta_3}$, delivering the local minimum of the objective function $\sigma^{\langle xy \rangle}$ within a defined range
\begin{equation}
\nonumber
\left[\left(\varepsilon,\alpha,\beta\right)^{a,b}_{\theta_1,\theta_2,\theta_3}\right]_{min}^{max}\,.
\end{equation}

$\left(\varepsilon,\alpha,\beta\right)^{a,b}_{\theta_1,\theta_2,\theta_3}$ are the variables the objective function $\sigma^{\langle xy \rangle}$ is being solved for. The objective function $\sigma^{\langle xy \rangle}$ to be minimized must be defined and some starting approximate values (initial estimate) have to be assigned to the variables $\left(\varepsilon,\alpha,\beta\right)^{a,b}_{\theta_1,\theta_2,\theta_3}$ before applying~$\Re^{\langle xy \rangle}$. The directly measured projected rms-emittance and Twiss parameters taken from Tab.~\ref{tab_5} have been defined naturally as the initial values. The function $\Re^{\langle xy \rangle}$ adjusts all the variables $\left(\varepsilon,\alpha,\beta\right)^{a,b}_{\theta_1,\theta_2,\theta_3}$ simultaneously in order to optimize the objective function $\sigma^{\langle xy \rangle}$ and take its smallest value.

Applying this numerical routine, all the standard deviations of the reconstructed coupled moments $\sigma^{\langle xy \rangle}$, $\sigma^{\langle xy' \rangle}$, $\sigma^{\langle x'y \rangle}$, and $\sigma^{\langle x'y' \rangle}$ can be separately minimized. Finally, the total objective function $\sigma$ is defined as
\begin{equation}
\sigma:=\sigma^{\langle xy \rangle}/\mathrm{mm^2}\,+\,\sigma^{\langle xy' \rangle}/\mathrm{mm~mrad}\,+\,\sigma^{\langle x'y \rangle}/\mathrm{mm~mrad}\,+\,\sigma^{\langle x'y' \rangle}/\mathrm{mrad^2}\,.
\end{equation}

\begin{figure}[hbt]
\centering
\includegraphics*[width=80mm,clip=]{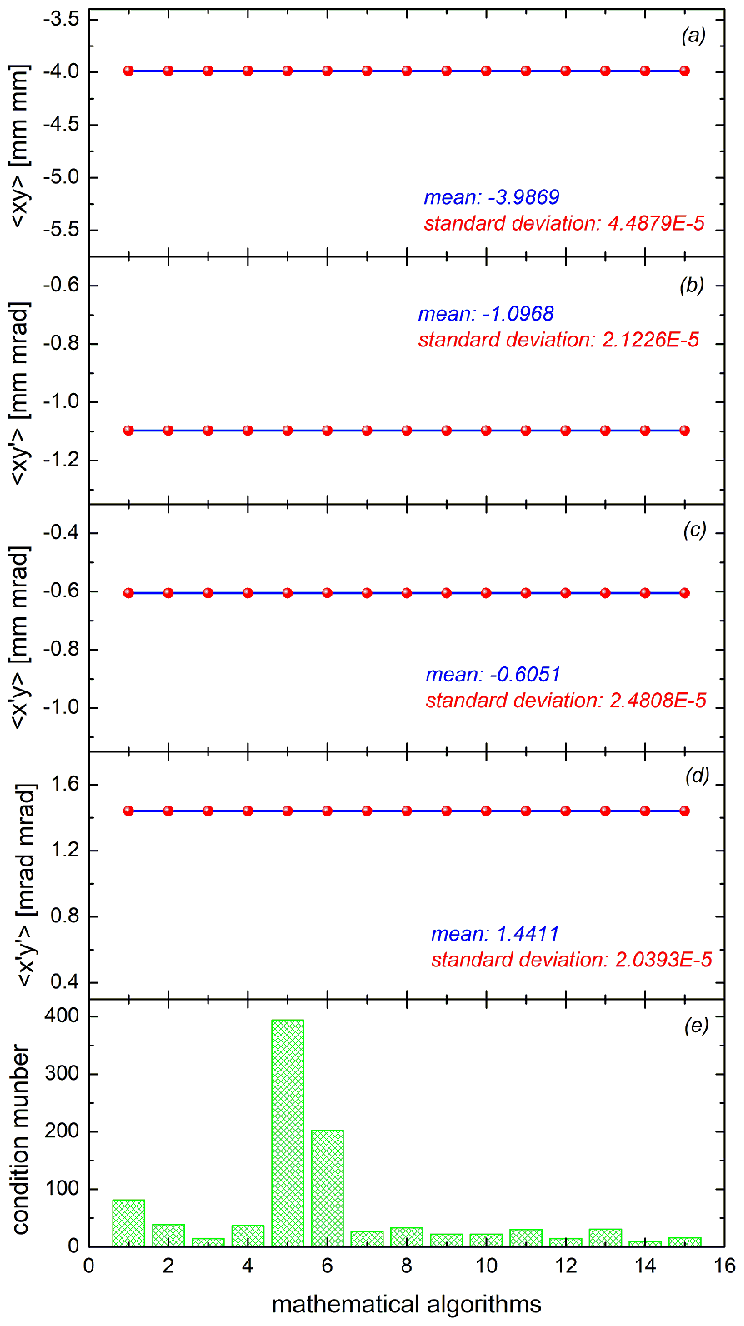}
\caption{Coupled moments obtained from trimmed measurements for the 15 possible selections. Dots indicate individual moments and solid lines indicate their means. (a) to (d): individual coupled beam moments and their means. (e): condition numbers.}
\label{correlated_moments_15_optimized}
\end{figure}

The value of total objective function $\sigma$ has been reduced from 4.546 to~0 successfully, and the trimmed projected Twiss parameters derived from the trimmed moments after optimization are listed in Tab.~\ref{tab_6}.
\begin{table}
\caption{\label{tab_6} Trimmed projected rms-emittances and Twiss parameters at rotation angles $\theta_1$=0$^{\circ}$, $\theta_2$=30$^{\circ}$, and $\theta_3$=90$^{\circ}$ using settings $a$ and $b$.}
\centering
\begin{tabular}{c|c|c|c|c}
\hline
\hline
Rotation& setting &$\alpha_{rms}$ & $\beta_{rms}$ [m/rad]& $\varepsilon_{rms}$ [mm~mrad]\\
\hline
0$^{\circ}$ & a &-0.135& 4.536 & 3.087\\
0$^{\circ}$ & b &0.100 & 3.954 & 3.106\\
30$^{\circ}$& a &-0.543 &2.263 &3.232\\
30$^{\circ}$& b &-0.801 &2.645 &4.697\\
90$^{\circ}$& a &-2.471 &8.705 & 3.395\\
90$^{\circ}$& b &-2.687 &7.374 & 3.333\\
\hline
\hline
\end{tabular}
\end{table}
Comparison of the trimmed Twiss parameters (Tab.~\ref{tab_6}) with the originally measured ones (Tab.~\ref{tab_5}) reveals that the trimming is below~2\% for each of the 18 trimmed measured values. The reconstructed coupled moments calculated from the trimmed measurements are shown in Fig.~\ref{correlated_moments_15_optimized}. The remaining standard deviation is practically zero, i.e., the coupled moments calculated from the 15 selections are consistent to each other and to the trimmed input. The amount of trimming (below~2\%) is less than the expected uncertainty of the measurements. Therefore, the averaged beam moment matrix from all 15 selections (in units of mm and mrad) applying trimmed 18 second moment measurements is 
\begin{equation}
\label{C_chen2}
C^\dagger=
\left(
\begin{array}{clcr}
+8.3830 &  -4.1962 & -3.9869   & -1.0968 \\
-4.1962  &  +3.2372 &  -0.6051  & +1.4411 \\
-3.9869   &  -0.6051   & +11.6647 & -3.1678\\
-1.0968   & +1.4411 &  -3.1678  & +1.8483
\end{array}
\right)\,.
\end{equation}

\begin{figure}[hbt]
\centering
\includegraphics*[width=80mm,clip=]{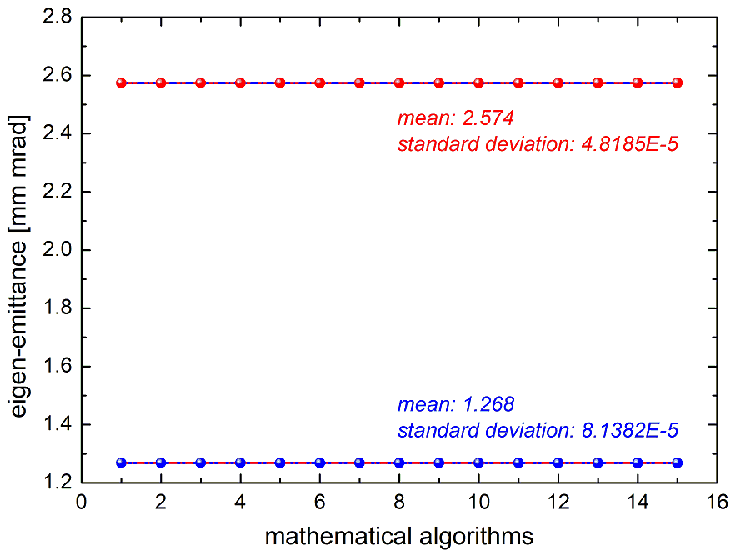}
\caption{Eigen emittances obtained from trimmed measurements for the 15 possible selections. Dots indicate individual eigen emittances and solid lines indicate their means. Dashed lines indicate the eigen emittances obtained from matrix $C^\dagger$.}
\label{eigen_emittances_2}
\end{figure}

Evaluation of the eigen emittances from matrix $C^\dagger$ results in $\varepsilon_1$=2.574~mm~mrad, $\varepsilon_2$=1.268~mm~mrad, and $t$=2.21. Figure~\ref{eigen_emittances_2} shows the eigen emittances calculated from the coupled moments shown in Fig.~\ref{correlated_moments_15_optimized}. The results for the eigen emittances are practically identical for all 15 selections. Comparing the spread, i.e. inconsistency, of eigen emittances shown in Fig.~\ref{eigen_emittances_1} (un-trimmed input) with the ones of Fig.~\ref{eigen_emittances_2} (trimmed input) reveals that just slight trimming of the input within the intrinsic uncertainty of the measurement delivers a most consistent set of input values and finally obtained results.

\begin{figure}[hbt]
\centering
\includegraphics*[width=80mm,clip=]{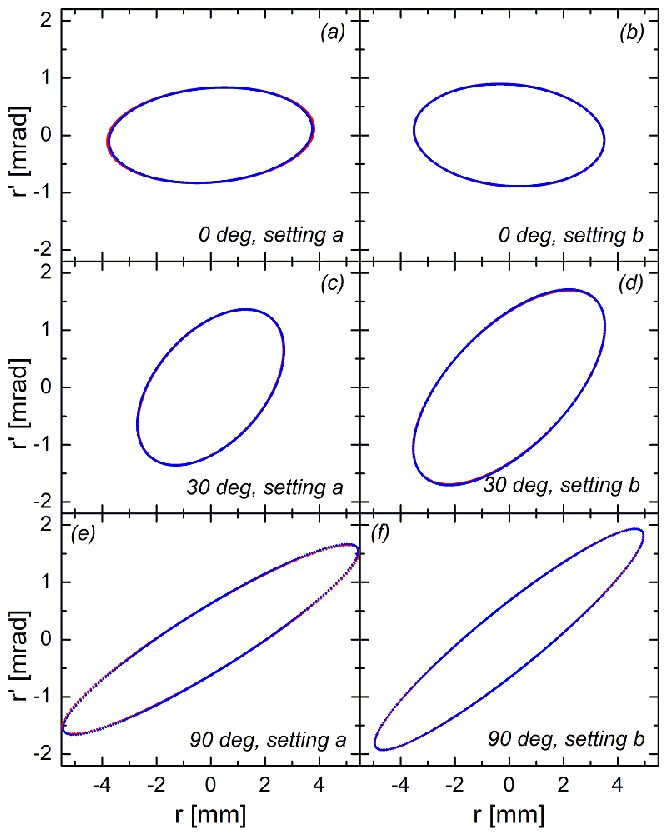}
\caption{Originally measured (red) and trimmed (blue) projected rms-ellipses at rotation angles $\theta_1$, $\theta_2$, and $\theta_3$ using settings $a$ and $b$.}
\label{measurement_and_optimization}
\end{figure}

Figure~\ref{measurement_and_optimization} illustrates how few trimming is needed to obtain a fully self-consistent set of 18 moments used to derive the desired coupled moments and eigen emittances. The figure plots the originally measured projected rms-ellipses and the corresponding trimmed rms-ellipses. They can be hardly distinguished.

It must be mentioned here that the applied method of trimming measured data such that they deliver an accurate result, may at first sight look like if data are modified to deliver any result and hence that there is arbitrariness in this result. This is not the case here as the procedure enforces that the data is self-consistent in the sense of strictly fulfilling~Eq.~(\ref{coupling001}) to~Eq.~(\ref{coupling006}). It must be considered that the 18 directly measured moments are not independent from each other; they are rather coupled to each other. This feature of intrinsically self-coupled physical input is exploited here. It shall obviously not be applied to input data that are not self-coupled as for instance in case of evaluating a density from measurement of a mass and a volume.
\section{Proper determination of decoupling lattice}
This section evaluates the qualification of the obtained results for the coupled moments to provide for an appropriate beam line, expressed in transfer matrix $R$, to decouple the beam. To this end the decoupling lattice is constructed in two ways: firstly assuming as input the results from the previous evaluation method written in beam matrix $C$, i.e., $R(C)$; secondly assuming as input the results from the new evaluation method written in beam matrix $C^\dagger$, i.e., $R^\dagger(C^\dagger)$. 
It is obtained that (in units of~m and~rad)
\begin{equation}
\label{R_chen}
R=
\left(
\begin{array}{clcr}
+0.8498 &  -0.0005 & +0.2984 & +0.5469\\
+0.4659 &  +1.5550 &  +0.5493 & +0.5010 \\
-0.0053& +0.0250 &+0.8566 & -1.6860\\
-0.0204  &  -0.2422& -0.0309& +0.7715 \\
\end{array}
\right)\,,
\end{equation}
\begin{equation}
\label{R_chen1}
R^\dagger=
\left(
\begin{array}{clcr}
+0.6229 &  -0.1589 & -0.0448 & +0.0147\\
+0.3200 &  +1.5550 &  +0.0166 & -0.9454 \\
+0.0488& -0.1666 &+ 0.6974 & -1.6860\\
+0.3131 &  +0.3438&  +0.0014 & +1.1183 \\
\end{array}
\right)\,.
\end{equation}

\begin{figure}[hbt]
\centering
\includegraphics*[width=80mm,clip=]{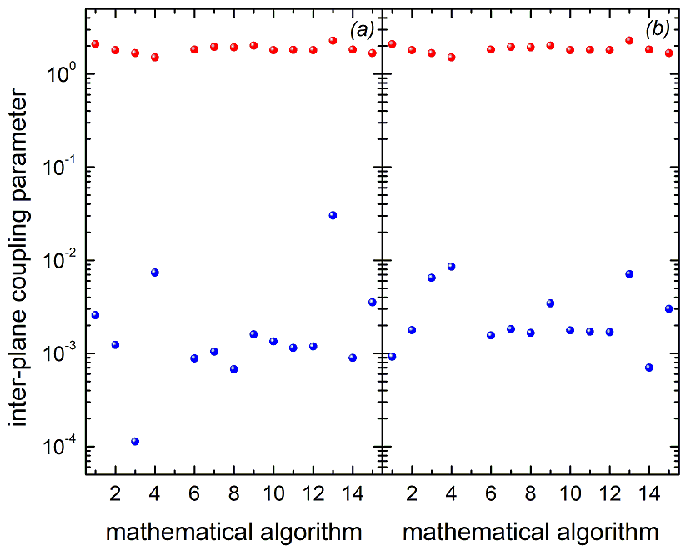}
\caption{Investigation of the decoupling capability of transfer matrices $R$ (left) and $R^\dagger$ (right). Red dots indicate $t(C_j)$, i.e, prior to decoupling. Blue dots indicate the residual coupling factors after decoupling. The data of selection \#5 are not plotted due to the unreliable evaluation result: before decoupling, $t\left(C_5\right)$=-1-0.795$i$~(imaginary).}
\label{decoupling}
\end{figure}

By construction evaluation of the coupling factor of the expressions $R C R^T$ and $R^\dagger C^\dagger {R^\dagger}^T$ delivers $t$=0.
In order to quantify the decoupling capability of both matrices with respect to the uncertainty in their determination, $R$ is applied to $C^\dagger$ and vice versa finding
\begin{equation}
\label{test111}
t\left(RC^\dagger R^T\right)=0.011~~~\mathrm{and}~~~t\left(R^\dagger C {R^\dagger}^T\right)=0.005\,.
\end{equation}
Accordingly, both evaluation methods will provide for a decoupling lattice that reduces the residual doupling practically to zero. Finally, one may evaluate
\begin{equation}
\label{test222}
t\left(RC_j R^T\right)~~~\mathrm{and}~~~t\left(R^\dagger C_j {R^\dagger}^T\right)\,,
\end{equation}
of which the result is plotted in Fig.~\ref{decoupling}. Residual coupling factors are lower than 0.1 in all cases and the beam is practically decoupled.
\section{Conclusions, Discussion and Outlook}
The ROSE device measures the full 4d beam matrix sufficiently accurate in order to construct a lattice that can decouple the beam. The method of evaluation its measurements has been improved such, that compared to the previous method, even the two eigen emittances are determined precisely. This was achieved by trimming the originally measured data, generally comprising a self-contradicting set of values due to intrinsic measurement inaccuracies, towards a fully self-consistent set of data that delivers very accurate results of the full 4d beam moment matrix.
\\
To our knowledge such a method has not been applied previously. A general mathematical base of this new trimming method and estimations for its limitations have not been worked out yet and are beyond the scope of this contribution. However, the results did not change even when enlarging the trimming range of the measured input data by a factor of two. It is subject of further research to form a more solid mathematical base for the method such that it can be easily adapted to other data evaluation problems including a systematic evaluation of the accuracy of the final result.
\bibliographystyle{model1a-num-names}











\end{document}